\documentclass[a4paper,fleqn,usenatbib]{mnras}

\usepackage{newtxtext,newtxmath}

\usepackage[T1]{fontenc}
\usepackage{ae,aecompl}

\usepackage{url}
\usepackage{pdflscape}
\usepackage{graphicx}	
\usepackage{amsmath}	
\usepackage{amssymb}	

\title[A Color-Excess Extinction map of the Southern Galactic disk from the VVV and GLIMPSE Surveys]{A Color-Excess Extinction map of the Southern Galactic disk from the VVV and GLIMPSE Surveys}
\author[M. Soto et al.]{
M. Soto,$^{1}$\thanks{E-mail: msoto@uda.cl}
R. Barb\'a,$^{2}$ 
D. Minniti,$^{3,4,5}$ 
A. Kunder,$^{6}$
D. Majaess,$^{7,8}$
\newauthor
J.L. Nilo-Castell\'on,$^{2,9}$ 
J. Alonso-Garc\'ia,$^{10,4}$
G. Leone,$^{1}$
L. Morelli,$^{1}$
L. Haikala,$^{1}$
\newauthor
V. Firpo,$^{2,11}$ 
P. Lucas,$^{12}$ 
J.P. Emerson,$^{13}$ 
C. Moni Bidin,$^{14}$ 
D. Geisler,$^{2,9,15}$ 
\newauthor
R.K. Saito,$^{16}$ 
S. Gurovich,$^{17}$
R. Contreras Ramos,$^{4,18}$ 
M. Rejkuba,$^{19}$ 
M. Barbieri,$^{1}$
\newauthor
A. Roman-Lopes,$^{2}$ 
M. Hempel,$^{3,18}$ 
M. V. Alonso,$^{17,20}$
L. D. Baravalle$^{17}$
\newauthor
J. Borissova,$^{4,21}$
R. Kurtev,$^{4,21}$ 
F. Milla$^{2}$
\\
$^{1}$Instituto de Astronom\'ia y Ciencias Planetarias de Atacama, Universidad de Atacama, Copayapu 485, Copiap\'o, Chile\\
$^{2}$Departamento de F\'isica y Astronom\'ia, Universidad de La Serena, Avenida Juan Cisternas 1200, La Serena, Chile\\
$^{3}$Departamento de Ciencias F\'isicas, Universidad Andr\'es Bello, Campus La Casona, Fern\'andez Concha 700, Santiago, Chile\\
$^{4}$Millennium Institute of Astrophysics, Av. Vicu\~na Mackenna 4860, 782-0436 Macul, Santiago, Chile\\
$^{5}$Vatican Observatory, Vatican City State V-00120, Italy\\
$^{6}$Saint Martin's University, 5000 Abbey Way SE, Lacey, WA 98503 USA\\
$^{7}$Saint Mary's University, Halifax, Nova Scotia, Canada\\
$^{8}$Mount Saint Vincent University, Halifax, Nova Scotia, Canada\\
$^{9}$Instituto de Investigaci\'on Multidisciplinario en Ciencia y Tecnolog\'ia, Universidad de La Serena. Benavente 980,La Serena, Chile\\
$^{10}$Centro de Astronom\'{i}a (CITEVA), Universidad de Antofagasta, Av. Angamos 601, Antofagasta, Chile\\   
$^{11}$Gemini Observatory, Southern Operations Center, La Serena, Chile\\           
$^{12}$Centre for Astrophysics Research, Science and Technology
             Research Institute, University of Hertfordshire, Hatfield
             AL10 9AB, UK\\
$^{13}$Astronomy Unit, School of Physics and Astronomy, Queen
             Mary University of London, Mile End Road, London, E1 4NS, UK\\
$^{14}$Instituto de Astronom\'{\i}a, Universidad Cat\'olica del Norte, Av. Angamos 0610, Antofagasta, Chile\\
$^{15}$Departmento de Astronom\'ia, Universidad de Concepci\'on, Casilla 160-C, Concepci\'on, Chile\\
$^{16}$Departamento de  F\'{i}sica, Universidade  Federal de Santa Catarina, Trindade 88040-900, Florian\'opolis, SC, Brazil\\
$^{17}$Instituto de Astronom\'ia Te\'orica y Experimental, (IATE-CONICET), Laprida 854, C\'ordoba, Argentina\\
$^{18}$Departamento de Astronom\'{\i}a y Astrof\'{\i}sica,
             Pontificia Universidad Cat\'olica de Chile, Vicu\~na Mackena
             4860, \\ 
             Casilla 306, Santiago 22, Chile\\
$^{19}$European Southern Observatory, Karl-Schwarszchild-Str. 2, D85748 Garching bei Muenchen, Germany\\
$^{20}$Observatorio Astron\'omico de C\'ordoba, Universidad Nacional de C\'ordoba, Laprida 854, C\'ordoba, Argentina\\     
$^{21}$ Departamento de F\'isica y Astronom\'ia, Facultad de Ciencias, Universidad de Valpara\'iso, Ave. Gran Breta\~na 1111, Playa Ancha,\\
Casilla 5030, Valpara\'iso, Chile      
}

\date{Accepted XXX. Received YYY; in original form ZZZ}

\pubyear{2018}

\begin{document}
\label{firstpage}
\pagerange{\pageref{firstpage}--\pageref{lastpage}}
\maketitle

\begin{abstract}
An improved high-resolution and deep A$_{Ks}$ foreground dust extinction map is
presented for the Galactic disk area within
$295^{\circ} \lesssim l \lesssim 350^{\circ}$, $-1.0^{\circ} \lesssim
b \lesssim +1.0^{\circ}$. At some longitudes the map
reaches up to $|b|\sim2.25^{\circ}$, for a total of $\sim$148 deg$^2$.
  The map was constructed via the Rayleigh-Jeans Color Excess (RJCE)
  technique based on deep near-infrared (NIR) and
  mid-infrared (MIR) photometry.  The new extinction map features a maximum
  bin size of 1$\arcmin$, and relies on NIR observations from the Two Micron All-Sky Survey (2MASS)
  and new data from ESO's Vista Variables in the V\'ia L\'actea (VVV) survey, in concert with MIR observations
  from the Galactic Legacy Infrared Mid-Plane Survey Extraordinaire (GLIMPSE).  The VVV photometry penetrates $\sim$4 magnitudes fainter than 2MASS, and provides enhanced sampling of the underlying stellar populations in this heavily obscured region.  Consequently, the new results supersede existing RJCE maps tied solely to brighter photometry, revealing a systematic underestimation of extinction in prior work that was based on shallower data. The new high-resolution and large-scale extinction map presented here is readily available to the community through a web query interface.
\end{abstract}



\newcommand{\arcdeg}{\(\stackrel{\:\circ}{\textstyle.\rule{0pt}{0.65ex}}\)} 

\section{Introduction}

Research pertaining to the structure and content of the Milky Way is paramount to 
understanding the broader universe. Yet a principal difficulty hindering such efforts is the presence of dense gas and dust along most sightlines.  Moreover, dust and gas are inhomogeneous and vary on small scales, thus propagating large uncertainties into low-resolution or magnitude-limited reddening maps, which are used to identify stellar populations sequences by enabling the determination of intrinsic colors. Those effects are exacerbated for sightlines toward the
Galactic plane and centre (Chen et al. 2013).


 Consequently, as a result of extinction in part there remains some uncertainty associated with  
 detailing the Galaxy's principal components, such as the disk, bulge, and halo. 
 Over the years considerable effort
 has been made to solve, or diminish as much as possible, these uncertainties by characterizing the interstellar dust distribution in Galactic extinction maps. Trumpler (1930) was one the first to try to determine the interstellar light absorption in several Galactic star clusters using spectroscopic techniques. The determination of color-excesses was based on the discrepancy between the expected color indices, based on stars in the solar neighborhood, and those actually observed. A major advancement in terms of the  reddening maps, is the work of Schlegel, Finkbeiner, \& Davis (1998; henceforth SFD), later superseded by Schlafly and Finkbeiner (2011), which produced an all-sky map based on 100 $\mu$m data from COBE/DIRBE and IRAS. Dust temperature was estimated using the ratio of 100 $\mu$m and 240 $\mu$m to trace the column density and derive the extinction by assuming a constant extinction law Rv=3.1. 
 However, the precision of the map rapidly decreases when approaching high extinction regions, such as the Galactic plane (Arce \& Goodman 1999, Gonzalez et al. 2012).
 This disadvantage limits the use of SFD in Galactic structure studies.   
 
 Since extinction is less pronounced at longer wavelengths maps based on infrared data have a comparative advantage over optical maps in the bulge and disk. Lada et al. (1994) assumed, based on data of the molecular cloud IC5146, an intrinsic color range of 0 $<$ (H-K)$_0$ $<$ 0.3 for a wide range of spectral types (A0 to M type stars) to obtain color excesses with an uncertainty of 2.5 magnitudes, in which was named the Near Infrared Color Excess method (NICE). 
 This technique was later modified by Lombardi \& Alves (2001) by using a combination of 2MASS (Skrutskie et al. 2006) near infrared colors (J-H) and (H-K$_{\rm S}$) in what they called the NICER (Near-Infrared Color Excess Revisited) method. The later technique was more recently improved by including a variable extinction law in the bulge (V-NICE; Gosling et al. 2009).  

Over the last decade,
considerable effort has been expended in pursuit of improving the large scale Galactic extinction
maps (e.g.  Gonzalez et al. 2011, 2012, 2018; Schultheis et al. 2014). 
Note that differential reddening will introduce
sizeable uncertainties into low-resolution and brightness-limited maps,
which we aim to help resolve. Indeed, we can obtain higher resolution through new surveys covering
large areas of the sky that provide an opportunity to improve upon
previous extinction maps, such as the Two Micron All Sky Survey
(2MASS; Skrutskie et al. 2006), the Sloan Digital Sky Survey (SDSS;
York et al. 2000), and the UKIRT Infrared Deep Sky Survey (UKIDSS;
Lucas et al 2008).  In this study, we will rely on NIR
observations from the ESO VISTA Variables in the V\'ia L\'actea (VVV)
public survey, in concert with MIR observations from the Galactic
Legacy Infrared Midplane Survey Extraordinaire (GLIMPSE; Benjamin 2003). GLIMPSE, 
 is a Legacy program that used the Infrared Array Camera (IRAC) on board of the \emph{Spitzer Space Telescope} to observe the Galactic Plane in 4 bands ($3.6$, $4.5$, $5.8$ and $8.0$) $\mu m$. 

The VVV Survey (Minniti et al. 2010) has among its principal objectives to bolster research of Galactic  structure. The
VVV survey is one of the initial six ESO VISTA public surveys, and has
sampled 562 square degrees of the Milky Way bulge (-10
$^{\circ}$$<$$\ell$$<$+10.25$^{\circ}$ and -10.25$^{\circ}$$< b
<$+5$^{\circ}$) and a section of the southern Galactic disk
(-65$^{\circ}$ $<$ $\ell$ $<$ -10$^{\circ}$ and -2.25$^{\circ}$ $< b <$
+2.25$^{\circ}$).  The survey has been carried out across five near-IR
bands ($Z$,$Y$,$J$,$H$,$K_{\rm s}$).  The final VVV observations
were carried out by October of 2015, completing a variability campaign of 6 years. 
 
%
%
 Thus, the area surveyed combined with the filter set used 
 ensure that the VVV has produced among the most
 detailed maps of the inner Galaxy in the near-infrared domain, providing
 important data needed to create enhanced extinction maps of our Galaxy.
 An example of this is the recent discovery of the low extinction windows in the VVV area (Minniti et al. 2018; Gonzalez et al. 2018). 

This study is organized as follows: the data employed to construct the extinction maps are characterized in Section 2, while the Rayleigh-Jeans Color excess method and the construction of our catalogues are described in Section 3. The constructed extinction map and a comparison with previous work are detailed in Section 4. Lastly, our findings are summarized in Section 5.

\section{Observations}
VVV observations were carried out using the \emph{Visible and
   Infrared Survey Telescope for Astronomy} (VISTA), located at the
 ESO Cerro Paranal Observatory.  VISTA is a 4.1 m telescope that is
 equipped with an Infrared Camera (VIRCAM; Emerson et al. 2006; Sutherland et al. 2015), which
 features an array of 16 2048 $\times$ 2048 pixels$^2$ Raytheon VIRGO
 detectors, and a pixel scale of $0."339$. The
 VIRCAM has a corrected 1.65 $deg$ diameter field of view,
 with its detectors arranged in a sparse 4$\times$4 array.  There is significant spacing between
 the detectors, corresponding to $42.5\%$ and $90\%$ of the array
 size along the x and y axis, respectively.  Each VISTA pointing is
 called a \emph{pawprint}, and covers  0.6 $deg^2$.  
 For gap-free sky coverage, a set of six offset pointings (known as pawprints) gives one filled rectangular tile. 
 One tile consists of a rectangle 1.475 $\times$ 1.017 $deg^2$, with each pixel covered by at least two of the six pawprints; plus two thin stripes each 0.092 $deg$ wide (along the two long edges) covered by one pawprint. In practice, each pawprint is usually comprised of several offset jitter positions for optimal removal of detector artefacts in later processing (Sutherland et al. 2015).  

 There is overlap between adjacent tiles, 
 within the 562 $deg^2$ of the complete survey 42 $deg^2$ are regions where tiles overlapped.

The VVV survey began in 2010, and is primarily a variability survey
with a baseline of over 5 years. The first year's observations
(2010-2011) were carried out in 5 broad-band filters (Z,Y,J,H,K$_{\rm
  s}$), while only K$_{\rm s}$ observations were planned for the
subsequent years (Minniti et al. 2010). 
Surveys (e.g., VVV) often require dedicated pipelines to reduce the amount of nightly data. The abundance of data associated with NIR surveys, relative to optical campaigns, stems from the need to overcome the instability of IR detectors and the sky brightness compared to stellar sources (Lewis et al. 2010).  Furthermore, IR sky emission varies over short timescales, with changes in spatial scale that can be significant. Thus, numerous short exposures are typically obtained to minimise such effects. 

The VVV data were reduced using the VISTA data flow system (VDFS) pipeline, which is running at the Cambridge Astronomical Survey Unit (CASU).  A detailed account of the CASU pipeline is described in Irwin et al. (2004) and Saito et al. (2012).  Briefly, the reduction process consists of a dark subtraction, reset correction, sky background subtraction, destriping correction of a low-level horizontal stripe pattern due to readout noise, jitter stacking of two slightly shifted pawprints to produce a \emph{stacked} pawprint, and tile construction whereby six stacked pawprints are combined.  
Single-band photometric catalogues are produced for
the stacked pawprints and tile images in each field, with
zeropoint calibrations as described by  Gonzalez-Fernandez et al. (2018).
The results presented in this work are based on the disk tiles located at $-65^{\circ} \lesssim l \lesssim-10^{\circ}$ and $-2.25^{\circ} \lesssim b \lesssim +2.25^{\circ}$.

MIR data was obtained through the IRSA (Infrared Science
Archive)\footnote{http://irsa.ipac.caltech.edu/cgi-bin/Gator/}, which is hosted by the NASA Infrared Processing
and Analysis Center (IPAC), and the observations correspond to the GLIMPSE I
(Benjamin et al. 2003) and GLIMPSE II/3D (Churchwell et al. 2009) 
survey areas overlapping with the VVV campaign. 

\section{Method}
 The Rayleigh-Jeans Color excess method (Majewski et al. 2011; henceforth M11) utilizes near-infrared and mid-infrared photometry, 
 since both sample the Rayleigh-Jeans part of the spectral energy distribution.  In that region, stars of diverse spectral types can share almost identical colors, which enables a precise determination of the extinction.
 
%
%
The main advantages of the RJCE method are: (1) it provides a
color-excess estimate for each star, and ensures an improved
dereddening solution using a CMD. By comparison, other techniques such
as NICE (Lada et al. 1994), NICER (Lombardi \& Alves 2001), V-NICE
 (Gosling et al. 2009), or the red clump method of Gonzalez et al. (2011), average or assume intrinsic colors for all stars analyzed. Lada et al. (1994) cites a maximum uncertainty for the calculated extinction A$_{\rm V}$ of 2.5 mag (NICE), as a consequence of the adopted average (H-K)$_0$ color. (2) The combination of NIR and MIR photometry removes a degeneracy that emerges for methods based solely on NIR colors, where the reddening vector and stellar evolution tracks run nearly parallel for most NIR color-color combinations. (3) Variations of the extinction law tend to be less significant for NIR colors (Indebetouw et al. 2005; Zasowski et al. 2009).

The RJCE color equation cited by Majewski et al. (2011) and Nidever et
al. (2012; henceforth N12) is restated below.
 \begin{equation}
  A(K_{\rm s}) = 0.918(H - [4.5]  - 0.08),
\end{equation}
where $H$ and $[4.5]$ correspond to a star's magnitude in the 2MASS and
GLIMPSE photometric systems.
The equation allows for the direct determination of the extinction per star and the construction of extinction maps over extended areas.

\subsection{Catalogues Construction}

To obtain a homogeneous sampling of the underlying stellar populations
in the three surveys, a combined catalogue was constructed following
the procedure described in detail by Soto et al. (2013). Briefly, the procedure starts 
by building the photometric transformations on a tile by tile basis that allows us to combine 
the VVV and 2MASS catalogues: i) we select the sources in the original CASU catalogues defined as stellar and clean them from detections in close proximity ($<$2$\farcs$0). ii) A small matching radius of 0$\farcs$3 is then used to combine the VVV resulting list with the 2MASS catalogue using STILTS (Taylor et
al. 2006), where only stars with Signal-to-noise ratio SNR $>$ 7 are used.
The resulting combined list VVV-2MASS of isolated stars provides the grounds to calculate linear fits between colors of the two catalogues that will be the photometric transformations. 
iii) Multiband VVV 
JHK$_{\rm s}$ for tile catalogues of all detections are generated with STILTS, where a maximum offset of 1$\farcs$0 to the nearest star is allowed during the
matching process.  
The derived VVV catalogue is then transformed to the 2MASS system using the photometric equations. It is in this VVV catalogue transformed to 2MASS photometric system that we replace with 2MASS measurements the magnitudes of stars brighter than K$_{\rm s} \sim 12.8\ mag$, typically saturated in the VVV disk catalogues (see also Gonzalez et al. 2012).
 iv) At this point we remove all multiple instances of sources from our VVV JHK$_{\rm s}$ catalogues as detected in the overlap region of the VVV tile edges.
A single datum is chosen based on a geometrical
criterion, where the source magnitude is selected from the tile furthest from the edge (see also 2MASS, Skrutskie et al. 2006).
 v) Finally, the resulting VVV-2MASS catalogue was then cross-referenced to 
 the GLIMPSE MIR photometry using a maximum matching radius of
 0$\farcs$5 to the nearest source.  
 The final map exhibits a total area of 148 \emph{sq. deg}.

\section{Results}
 
The merged 2MASS, VVV, and GLIMPSE stellar density map for the Galactic midplane is shown in Figure \ref{fig:Nmap_ext}. 
The number of stars per resolution element of 1' ranges from 1 to 66, with the median being 23.  
The stellar distribution is quite smooth spatially, with only a clear artefact seen as a diagonal strip at $l\sim328^{\circ}$ due to limited number of stars from the GLIMPSE survey. Errors for every pixel in the map were derived from the photometric uncertainties. 
 As expected, the regions closest to the Galactic centre ($350^{\circ} \lesssim l \lesssim 340^{\circ}$) tend to have the highest stellar density.  

Figure \ref{fig:lbmap_ext} 
displays the generated $A(K_{\rm S})$ maps.  The extinction decreases
precipitously away from the plane, although there exist some highly
obscured regions at higher latitudes.  Given the RJCE method measures
the reddening for each star, in concert with the high density along
the plane and the depth of the VVV and GLIMPSE surveys: the resulting reddening map is particularly interesting for studies of highly extincted regions. The extinction results can be
retrieved online at the project's webpage\footnote{\ http://www.astro.uda.cl/msoto/extinction/ab.php\\
or http://astro.userena.cl/ExtMapVVV/}.

\begin{landscape}
  \begin{figure}
   \centering
   \includegraphics[width=23cm]{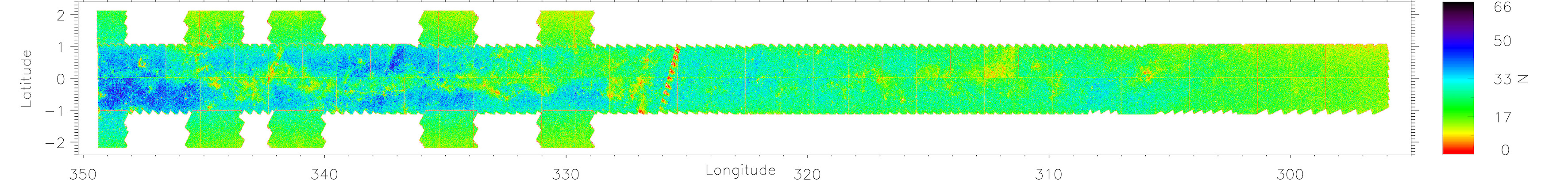}
   \caption{Star count map (1'$\times$1' bin size) for common fields
     between the VVV and GLIMPSE surveys of the Southern Galactic
     disk.
}
              \label{fig:Nmap_ext}%
    \end{figure}

  \begin{figure}
   \centering
   \includegraphics[width=23cm]{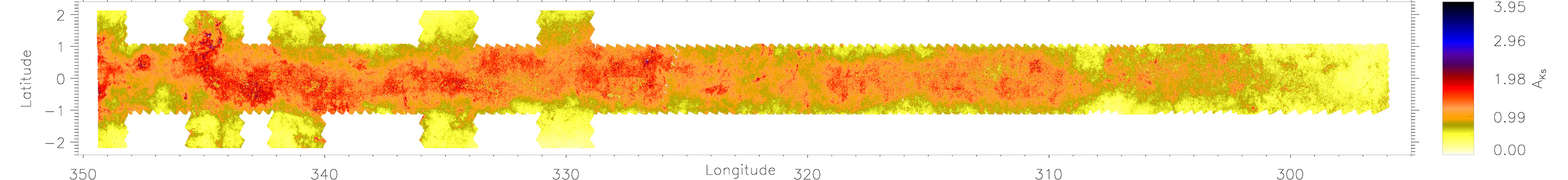}
   \caption{New improved extinction map constructed by applying the Rayleigh-Jeans Color excess method to the region in Fig.~\ref{fig:Nmap_ext}.  
}
              \label{fig:lbmap_ext}%
    \end{figure}

\begin{figure}
   \centering
   \includegraphics[width=10cm]{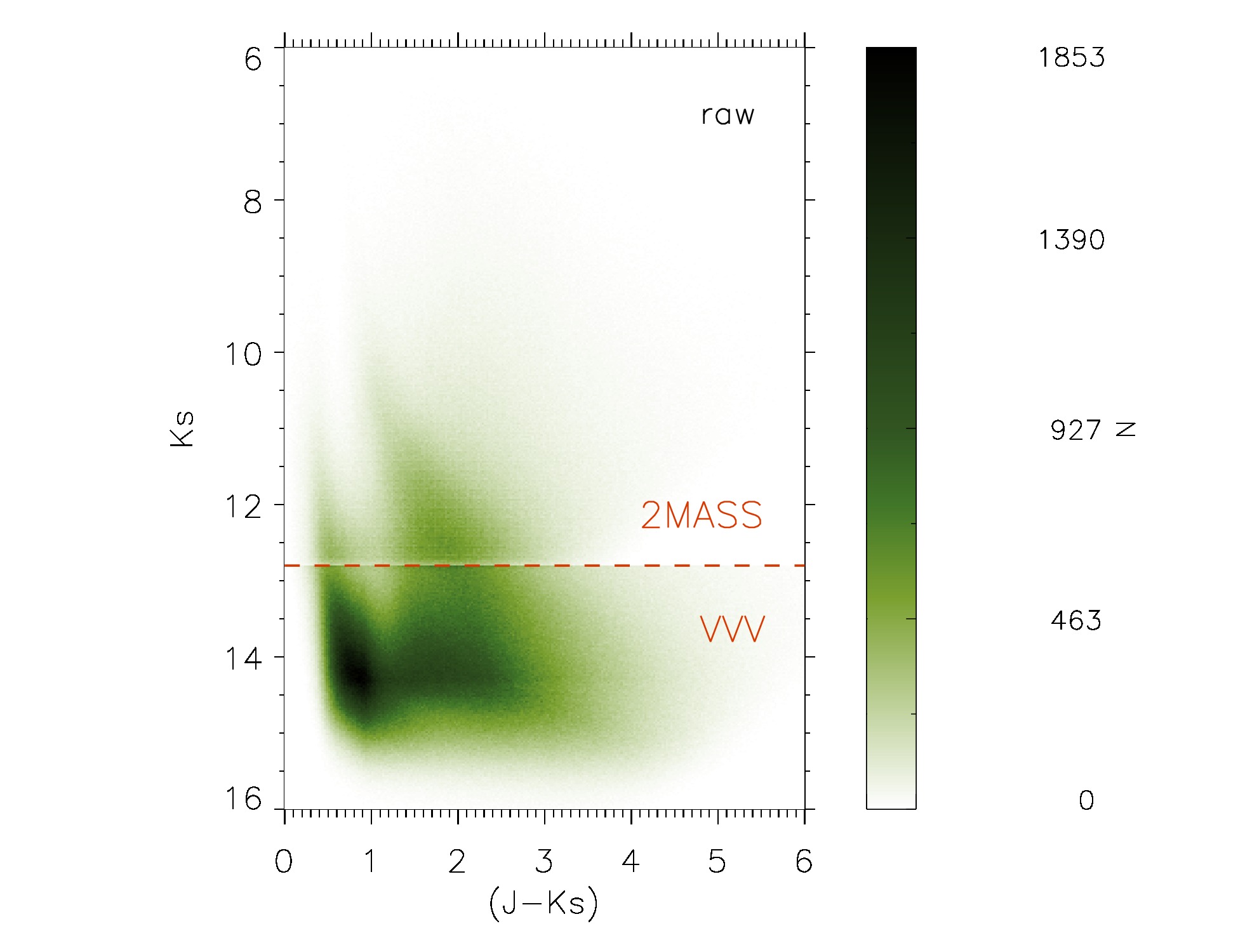}
   \includegraphics[width=10cm]{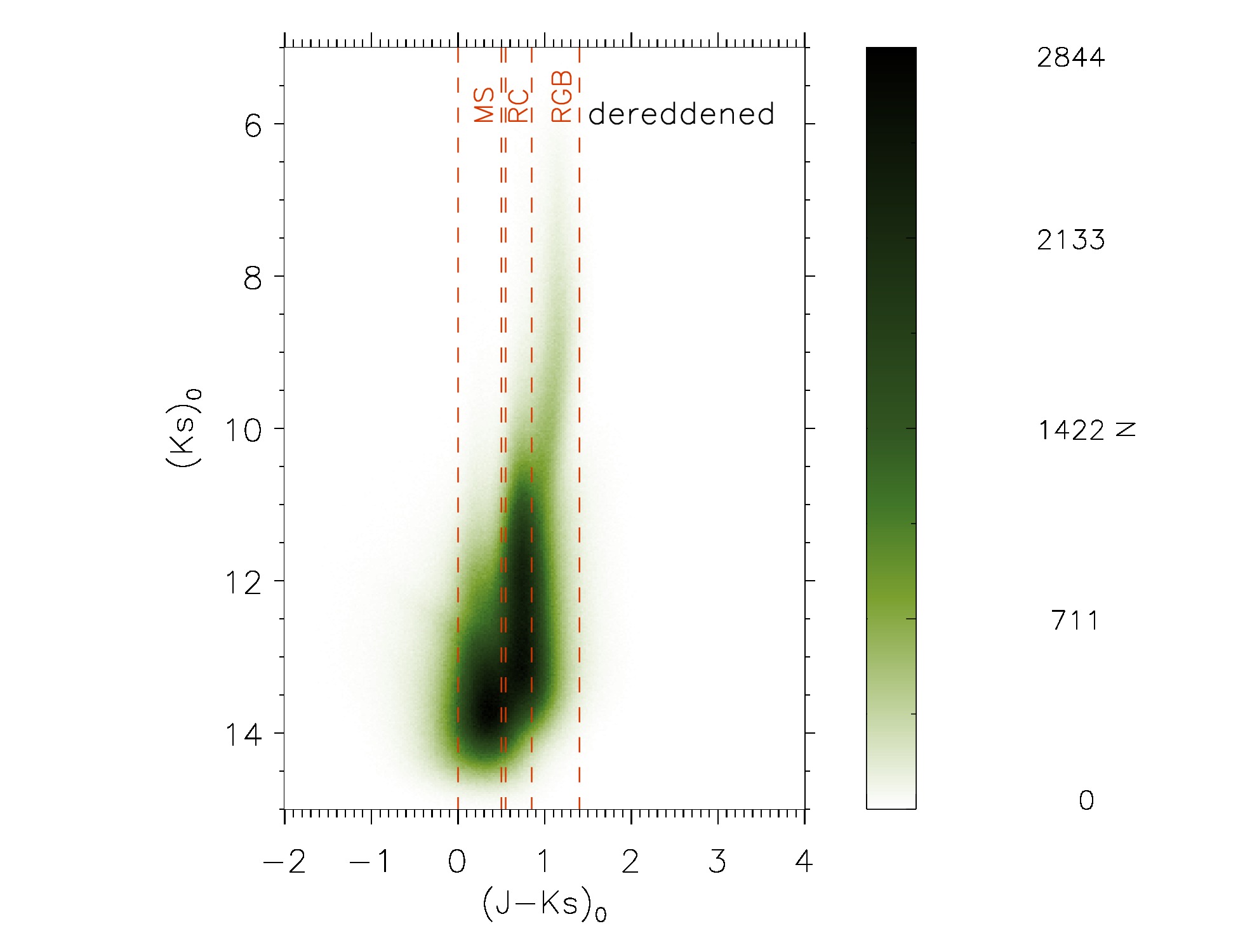}
   \caption{Hess diagrams 
    corresponding to the area of the maps of Figures \ref{fig:Nmap_ext} and \ref{fig:lbmap_ext} and 240$\times$400 bins.
   \emph{Left}, raw CMD,  where the VVV
   magnitudes were transformed to the 2MASS photometric system. Stars with 
  $K_{\rm s} <$12.8 (red dashed line) were replaced with photometry from the 2MASS point source catalogue in order to avoid saturation in
  the VVV catalogues. \emph{Right}, dereddened CMD for the same 
 stars, where the vertical lines highlight the same cuts used in Nidever et
 al. (2012) to divide stellar evolutionary phases..
}
              \label{fig:cmd_all_ext}%
\end{figure}
\end{landscape}

\begin{landscape}
\begin{figure}
   \centering
   \includegraphics[width=10cm]{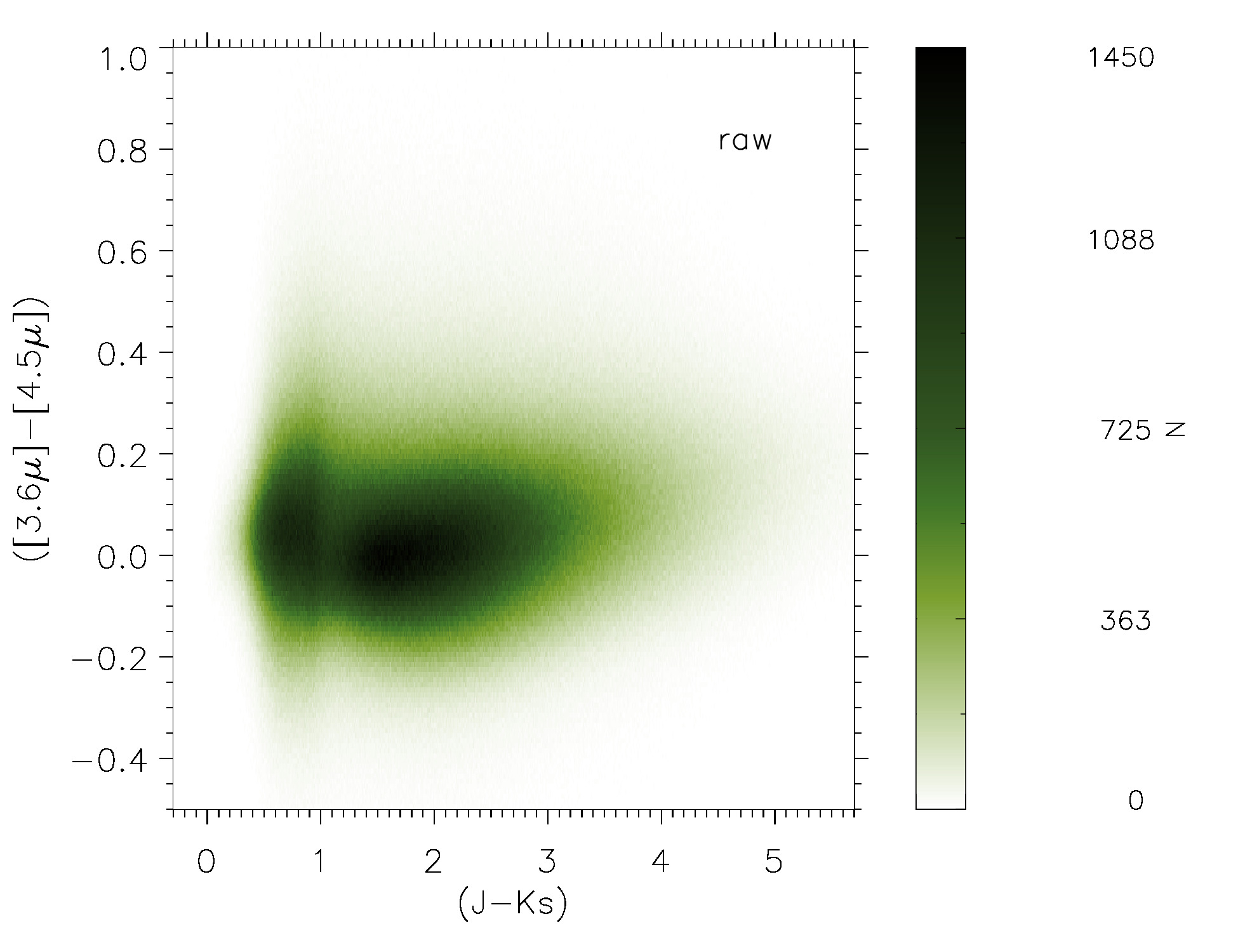}
   \includegraphics[width=10cm]{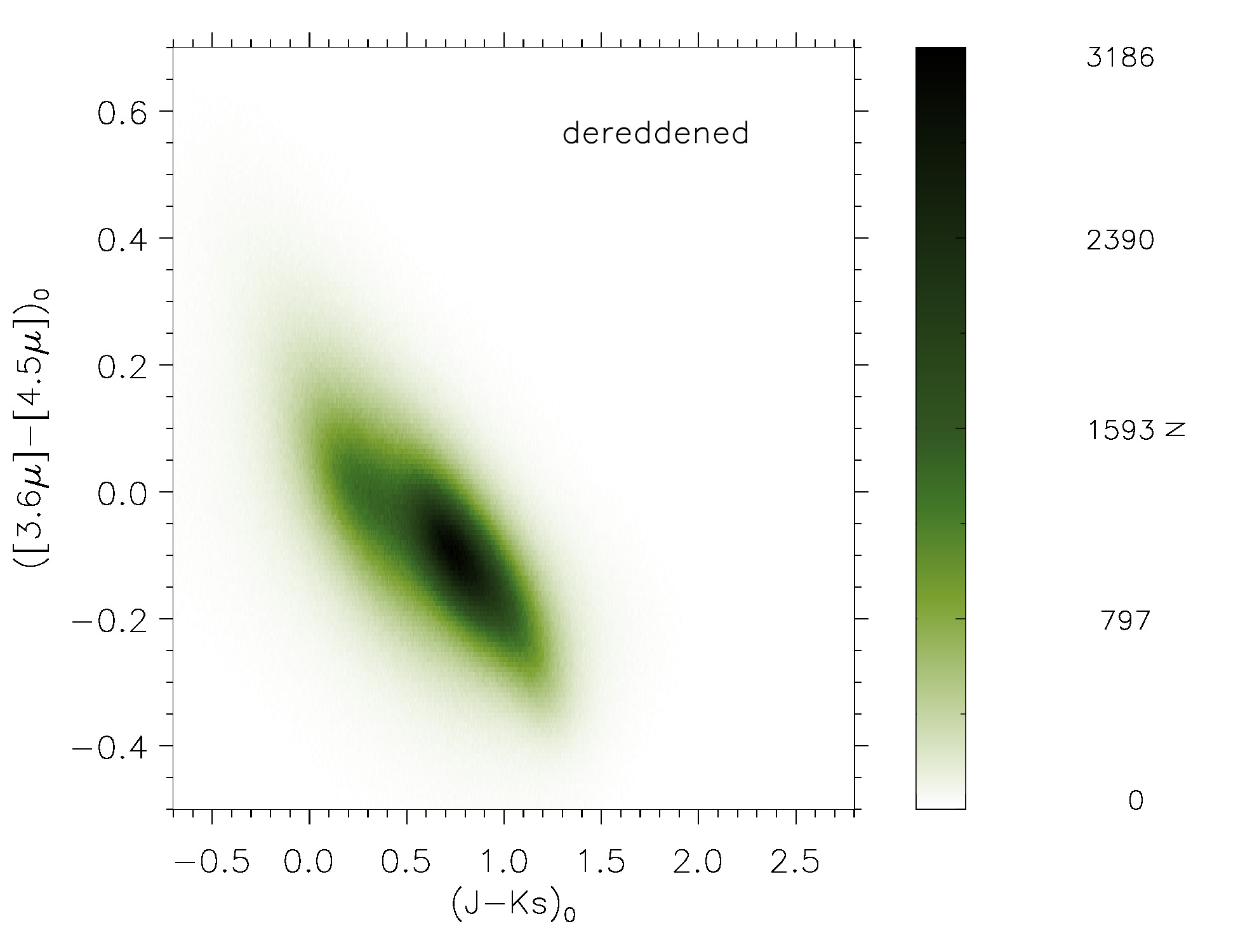}
   \caption{Same as figure \ref{fig:cmd_all_ext}, but for the color-color diagrams in (J-K$_{\rm s}$) and (3.5$\mu$-4.5$\mu$) colors. \emph{Left}, raw (reddened) color-color diagram, \emph{right} dereddened.}
              \label{fig:color_all_ext}%
\end{figure}

\end{landscape}

\begin{landscape}
  \begin{figure}
   \centering
  \includegraphics[width=21cm]{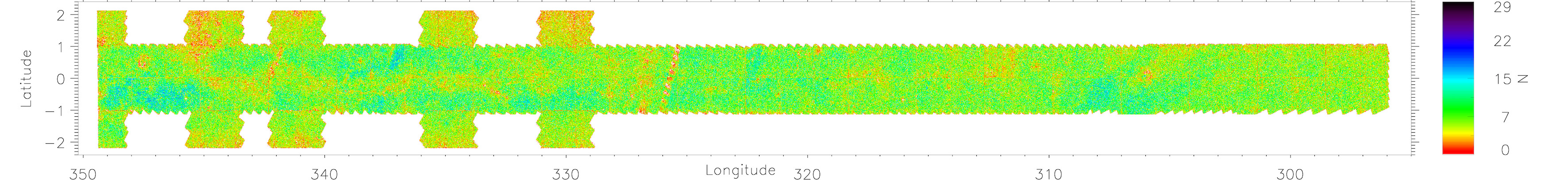}
   \includegraphics[width=21cm]{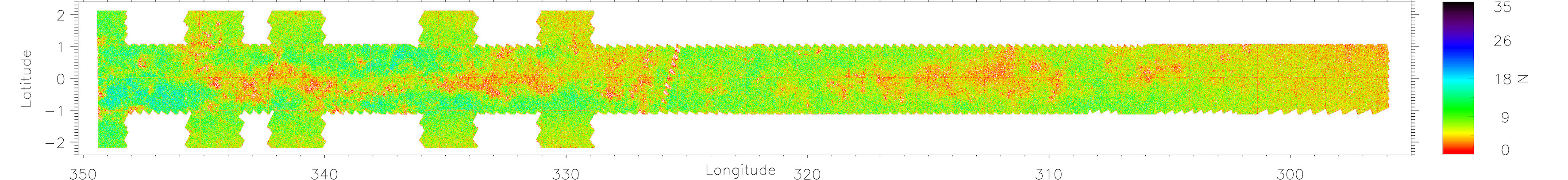}
   \includegraphics[width=21cm]{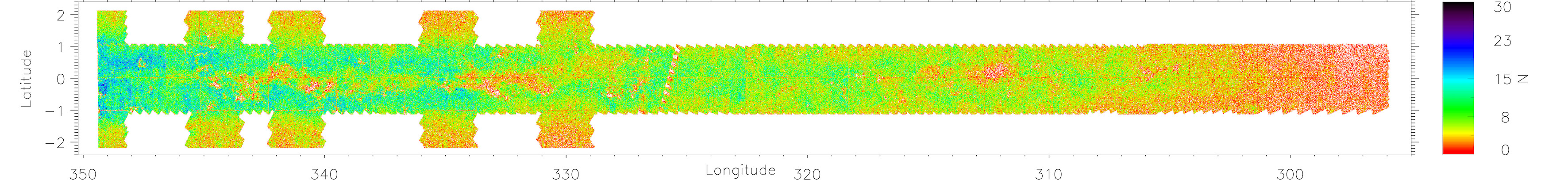}
   \caption{
     Same as Fig. \ref{fig:Nmap_ext}, but divided by stellar
     populations using the cuts of Figure \ref{fig:cmd_all_ext}. \emph{From top to bottom},
      star count map for the main sequence, red clump, and RGB, respectively.
}
              \label{fig:Nmap_ext2}%
    \end{figure}

  \begin{figure}
   \centering
   \includegraphics[width=21cm]{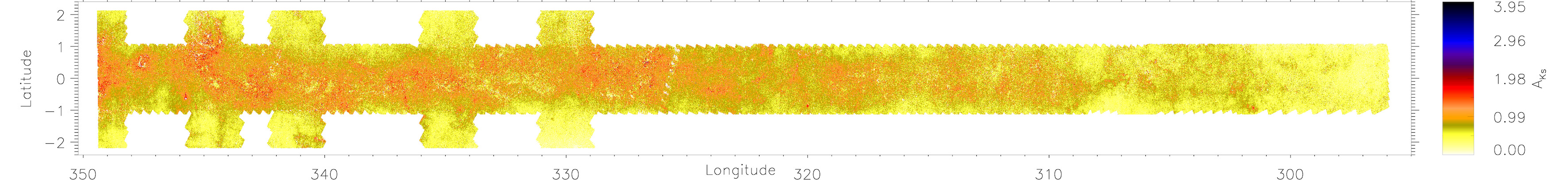}
   \includegraphics[width=21cm]{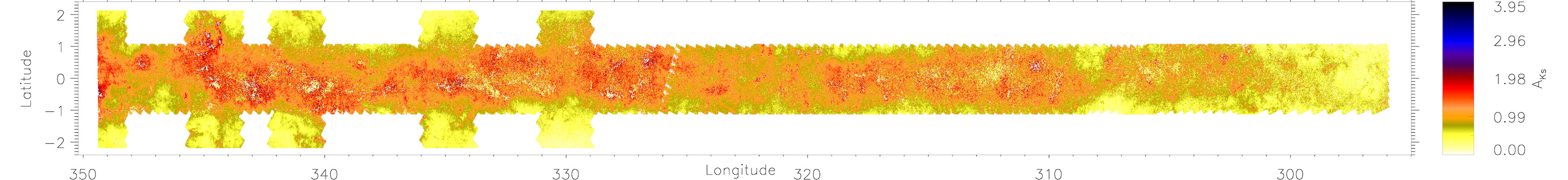}
   \includegraphics[width=21cm]{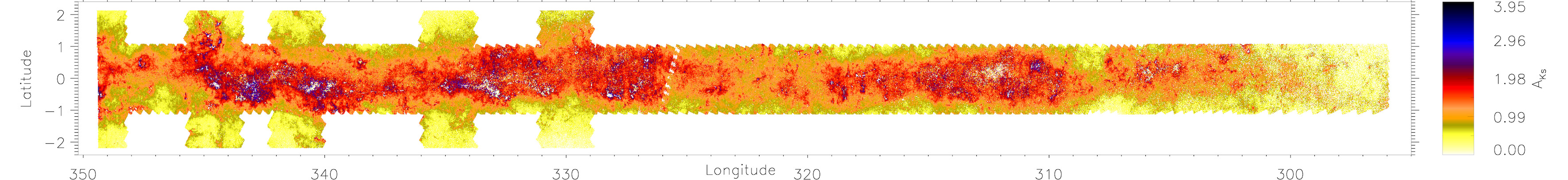}
   \caption{
    Same as Figure \ref{fig:lbmap_ext}, but discriminating by stellar
    populations using the cuts of Figure \ref{fig:cmd_all_ext}.
     \emph{From top to bottom}, extinction map for main sequence, red clump, and 
    RGB stars, respectively. 
}
              \label{fig:lbmap_ext2}%
    \end{figure}
\end{landscape}

Figures \ref{fig:cmd_all_ext} and \ref{fig:color_all_ext}
convey the RJCE results for the stellar population in the southern
 Galactic disk, where the VVV fields overlap with the GLIMPSE
 footprint.  The raw IR CMD and color-color diagram feature the expected dispersion of stars
 resulting from fields with significant differential extinction, such
 as those near the Galactic plane. 
 The dereddened CMD displays the same stars, but a narrower sequence is
 apparent and different stellar populations are distinguishable.  In
 particular, the Main Sequence (MS), Red Clump (RC), and Red Giant
 Branch (RGB) are separable using the $(J-K_{\rm S})_0$ color.
 The sequences are labelled according to $(J-K_{\rm S})_0$ boundaries defined in M11.
 The run of
 extinction as a function of distance can be computed, allowing clouds
 and broader dust complexes to be identified. 
 Similarly, the dereddened color-color diagram shows the expected decrease in the dispersion, in agreement with the histograms of selected color combinations shown in M11 and based on the Girardi et al. (2002) isochrone set.

Figure \ref{fig:Nmap_ext2} features the stellar density of the RGB, RC, and MS populations, whereas 
Figure \ref{fig:lbmap_ext2} displays the corresponding extinction map. 

\begin{figure}
   \includegraphics[width=8.8cm]{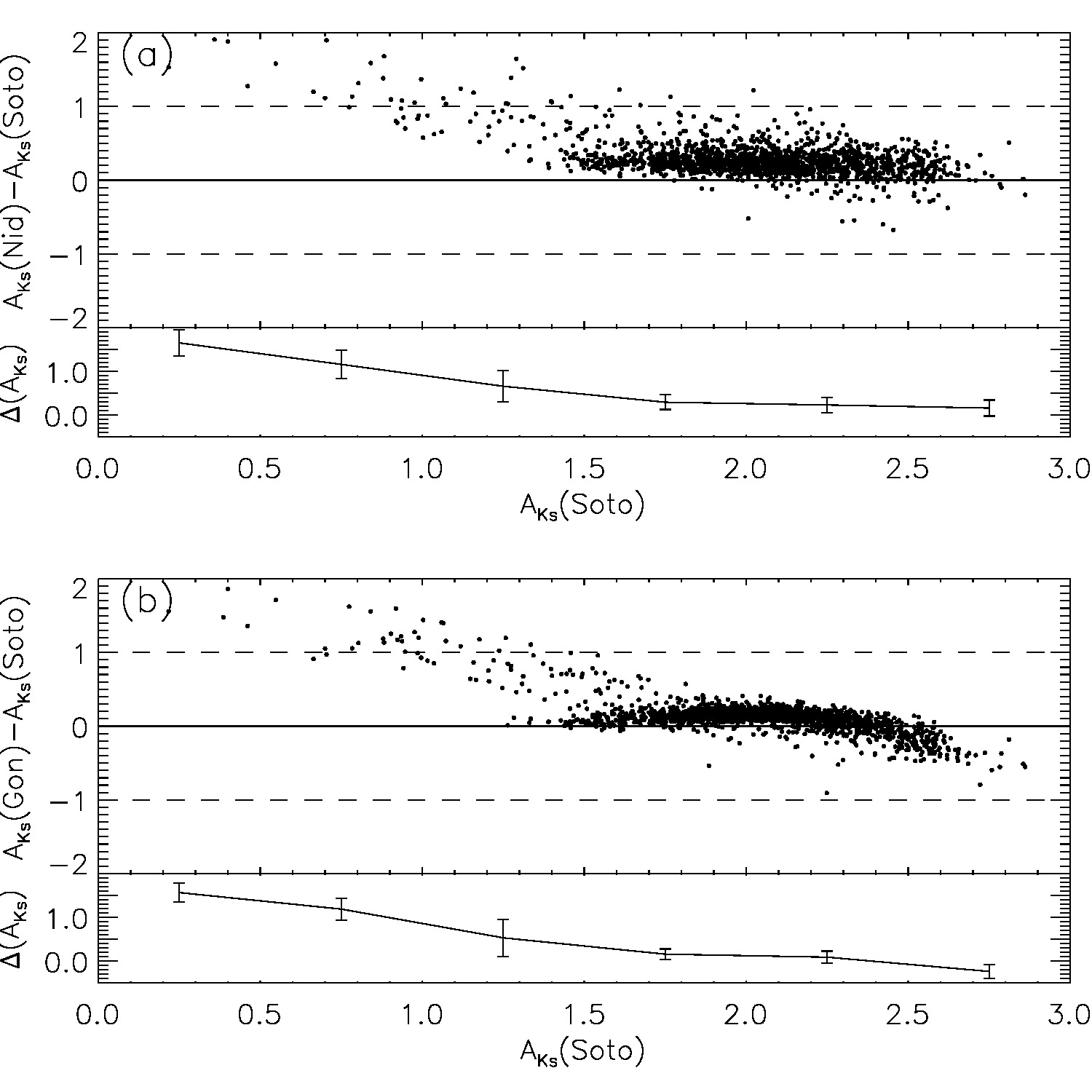}
   \caption{
     Direct comparison of the extinction per map-pixel in the VVV field
     b328 ($l$,$b$)=(7.9$^{\circ}$,0.05$^{\circ}$) derived from this
     work and compared with the results of:
  (\emph{a}) Nidever
  et al (2012), (\emph{b}) BEAM (Gonzalez et
  al. 2012). A pixel of
     2'$\times$2' has been used for consistency.  
}
              \label{fig:compare_ext}%
\end{figure}

A comparison of extinction maps made with the RJCE-VVV and other methods now follows and was selected for an area with data common across the different methods.
Figure \ref{fig:compare_ext} conveys a comparison with the G12 and N12 extinction maps, in the area corresponding to the VVV bulge field
\emph{b238} ($l,b=7.9^{\circ},0.05^{\circ}$). The N12 map is based on the RJCE technique applied to 2MASS data and
allows options for the population selection: main
sequence and turn-off (MSTO), RC, and RGB.
The G12 results are tied to a fit of the mean
($J-K_{\rm S}$) color for RC stars, calibrated using the colors of the
population endemic to Baade's window. The G12 results were obtained from 
 the BEAM 2 site\footnote{https://www.oagonzalez.net/beam-calculator} and translated to $A_{Ks}$ using $A_{Ks} = 0.640\times E_{(J-Ks)}$.
%
In order to enable a direct pixel-by-pixel comparison, the RJCE-VVV-GLIMPSE extinction map presented here was degraded to match the $2'$ resolution of the G12 and N12 maps.
Figure \ref{fig:compare_ext} relays the agreement with the N12 map,
 where the options ALL and PER were selected using Nidever's software\footnote{maps and software are currently available at:\\
https://www.noao.edu/noao/staff/dnidever/rjce/\\ 
extmaps/index.html} (corresponding to the 90th percentile of the total sample of stars irrespective of stellar population in the deredenned CMD). 
%
Similarly, the G12 
map is also in agreement with the results presented here. 
A difference exists for a small group of pixels exhibiting $A_{Ks}(Soto)<1.3$, whereas they appear with higher $A_{Ks}$ in N12 and G12. The difference likely stems from structure around $(l,b)\simeq(352^{\circ},-0.3^{\circ})$ in the field $b328$.
  A closer inspection indicates that those pixels with $A_{Ks}(Soto)<1.3$ in our map are statistically dominated by main sequence stars (58\% of them). The difference relative to G12 emerges because they rely on RC stars to determine extinction, while N12 is restricted by the brighter limit of 2MASS. 
 The 3D extinction map results by Schultheis et al. (2014; henceforth S14) were examined to discern if these low extinctions pixels are present elswhere. S14 produced a 3D extinction map based on VVV and 2MASS data, and by fitting the M giants temperature-color and distance color relations to the Besan\c{c}on model. The result is a 3D extinction map that has been integrated to several distances in the Galactic bulge and is available to the community\footnote{http://vizier.cfa.harvard.edu/viz-bin/\\
 VizieR?-source=J/A+A/566/A120}. 
 Similarly, we find a difference in our results with respect to N12 in the region $(l,b)\simeq(351.5^{\circ},0.65^{\circ})$, where a significant fraction of the population on those pixels 48\% (36/75) are MS or highly extincted stars below the faint limit of 2MASS. Thus, the differences found respect to N12 can be explained by the stars not used in N12 due to the limitations of their data.

\begin{figure*}
   \centering
   \includegraphics[width=8cm]{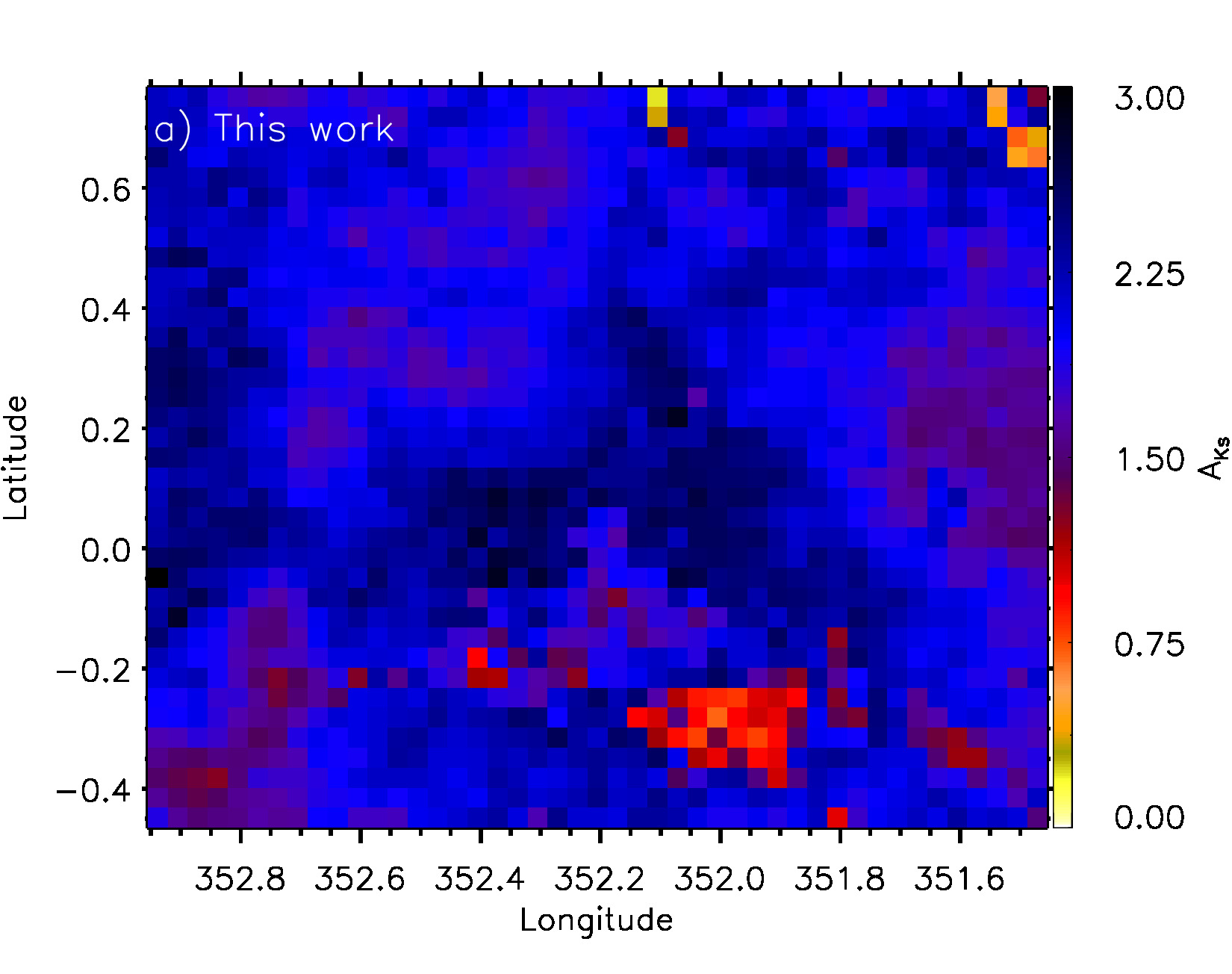}
   \includegraphics[width=8cm]{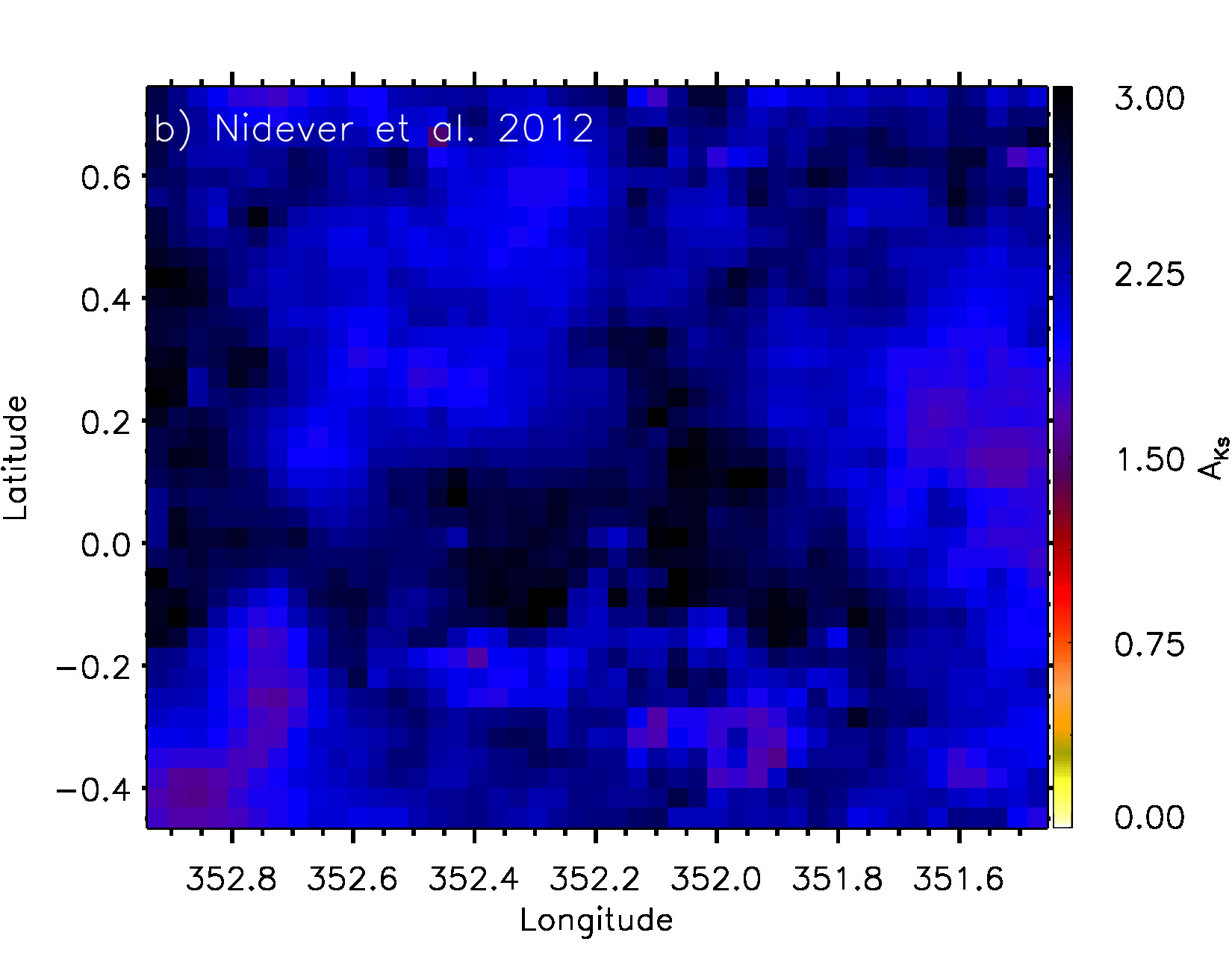}\\
   \includegraphics[width=8cm]{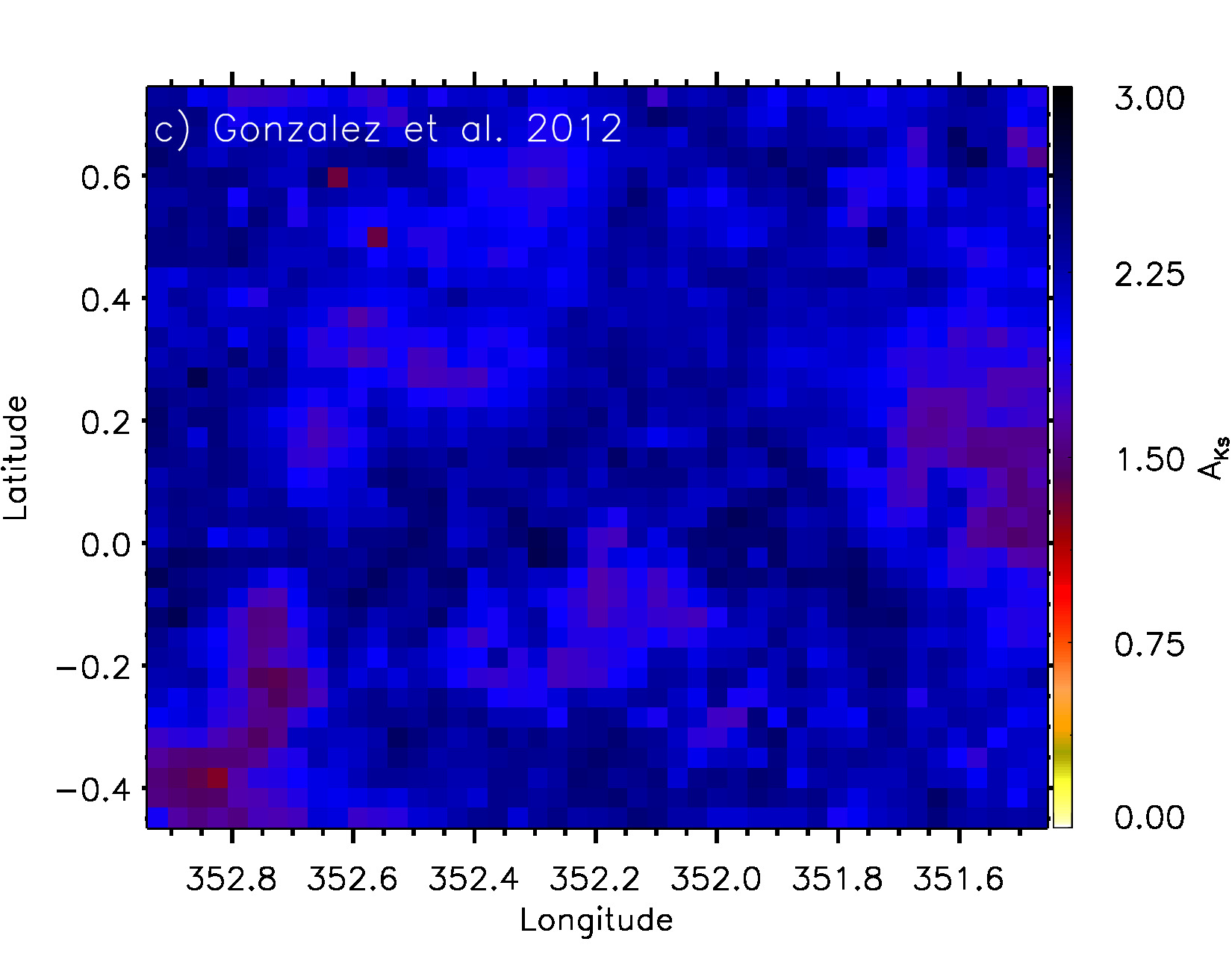}
   \includegraphics[width=8cm]{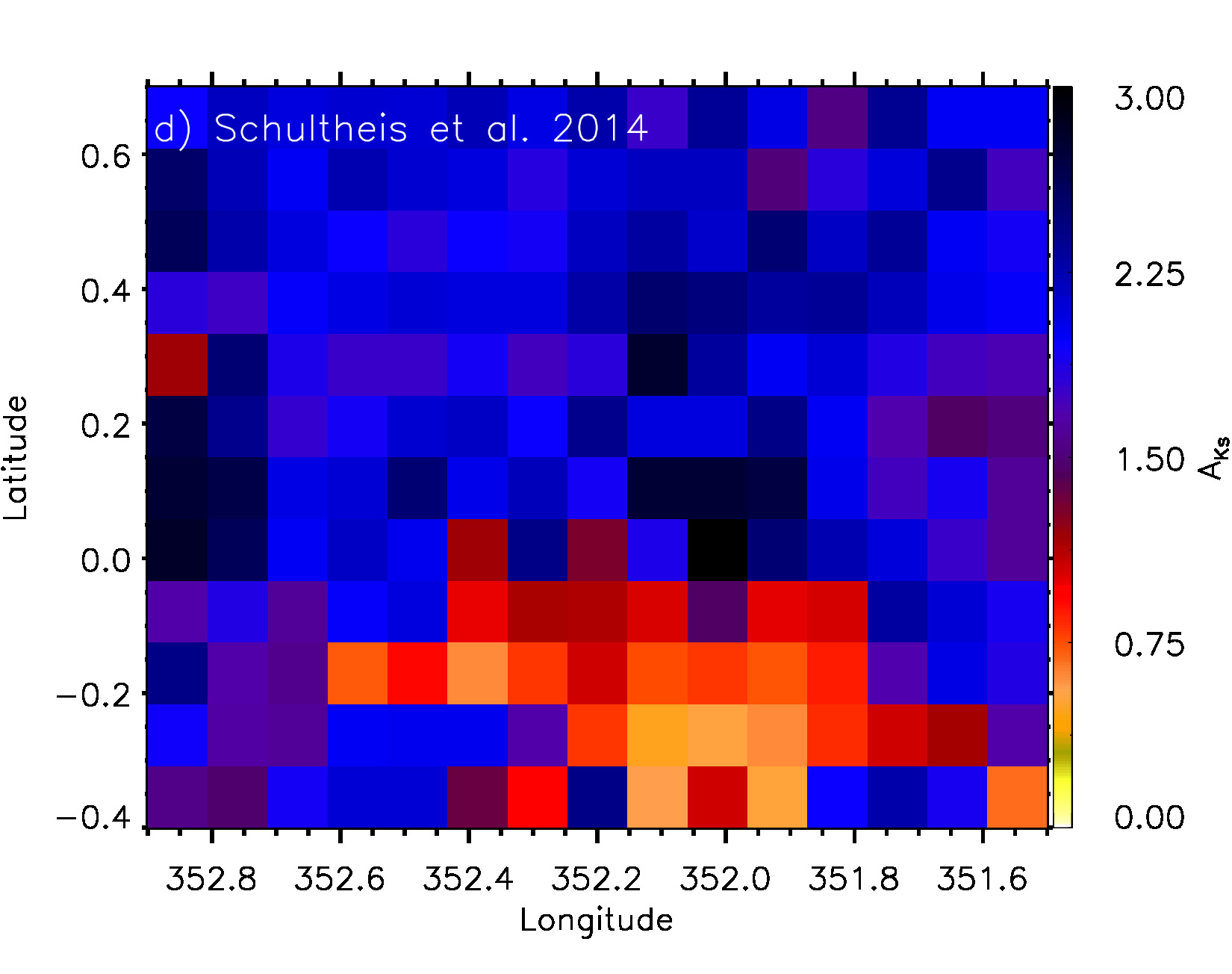}
   \caption{Extinction maps for VVV field \emph{b328}: \emph{a)} this work, 
   \emph{b)} Nidever et al. 2012, \emph{c)} Gonzalez et al. 2012, and
   \emph{d)} Schultheis et al. 2014.
}
              \label{fig:compare_ext_4}%
\end{figure*}

 Figure \ref{fig:compare_ext_4} shows the $A_{Ks}$ maps for the VVV field $b328$ in this work, G12, N12, and S14. For the latter, the map was constructed by using the reported $A_{Ks}$ values for an integrated distance of 8 $kpc$ and 0.1$^{\circ}$ grid. The corresponding $A_{Ks}$ maps agree, and in particular the same low extinction zone is present in the S14 and the present work's maps ($(l,b)\simeq(352^{\circ},-0.3^{\circ})$).

Other possible source behind the small discrepancies could be related to the
photometric color transformation equations used when developing the maps. This possibility may be explored by reconstructing the full map of the disk using only VVV data with deeper psf photometry in the future.

\begin{figure*}
   \centering
   \includegraphics[height=5.5cm]{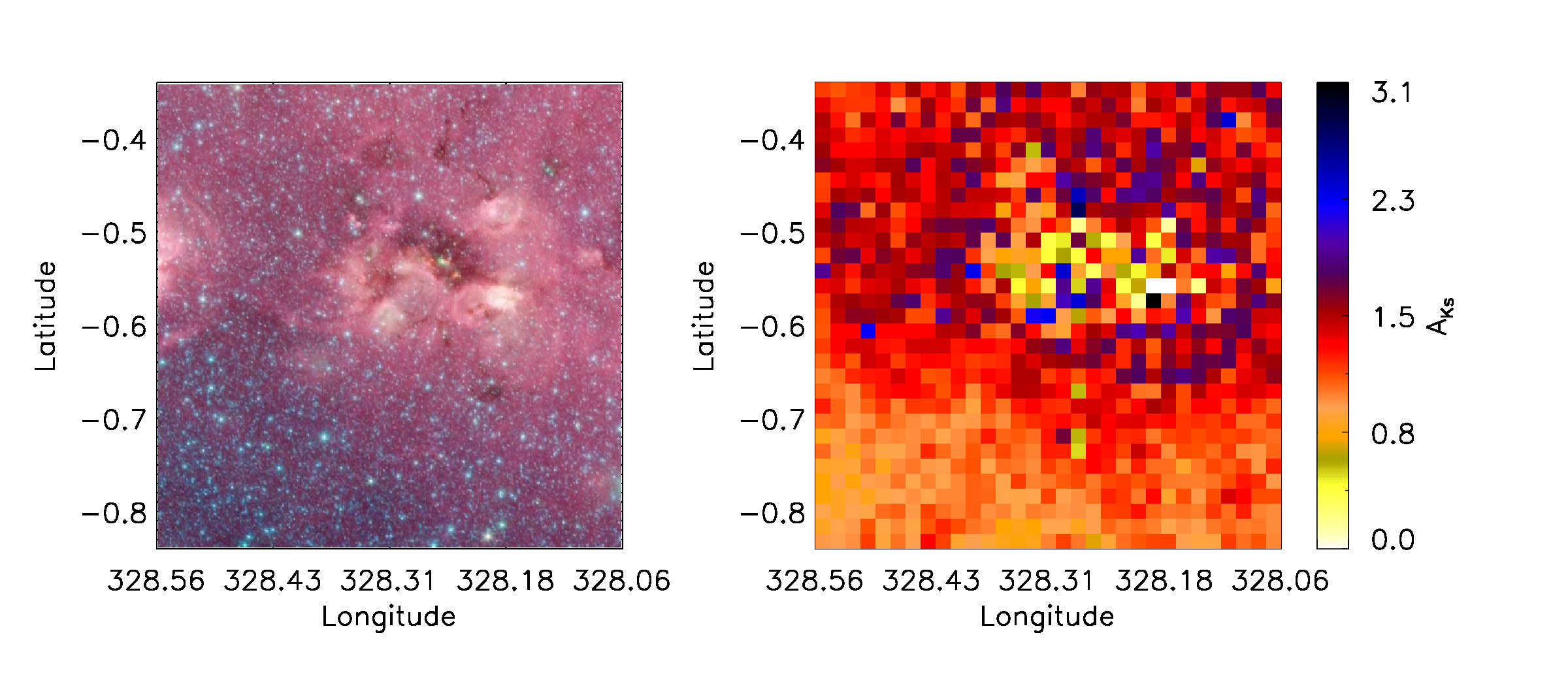}
   \includegraphics[height=5.5cm]{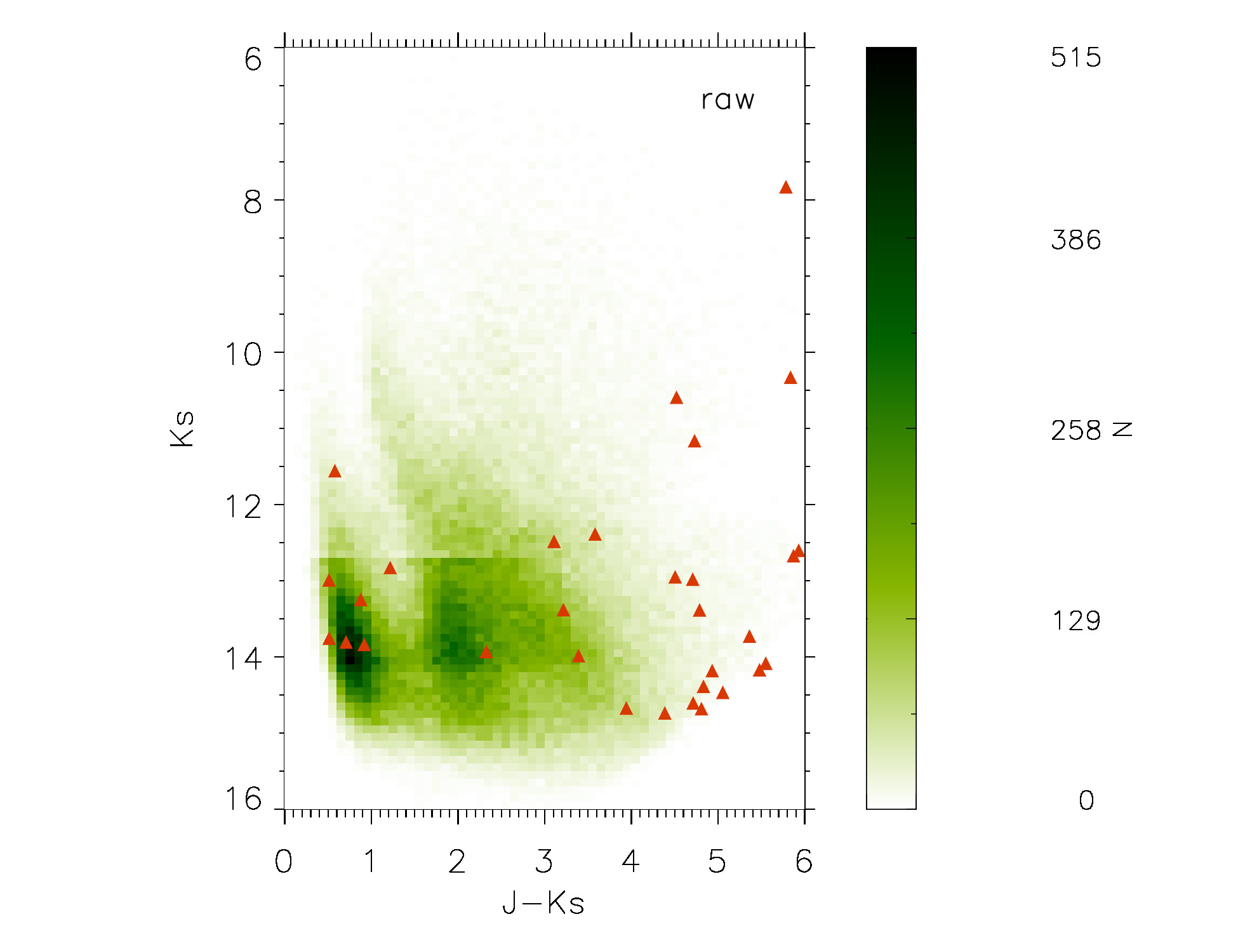}
   \includegraphics[height=5.5cm]{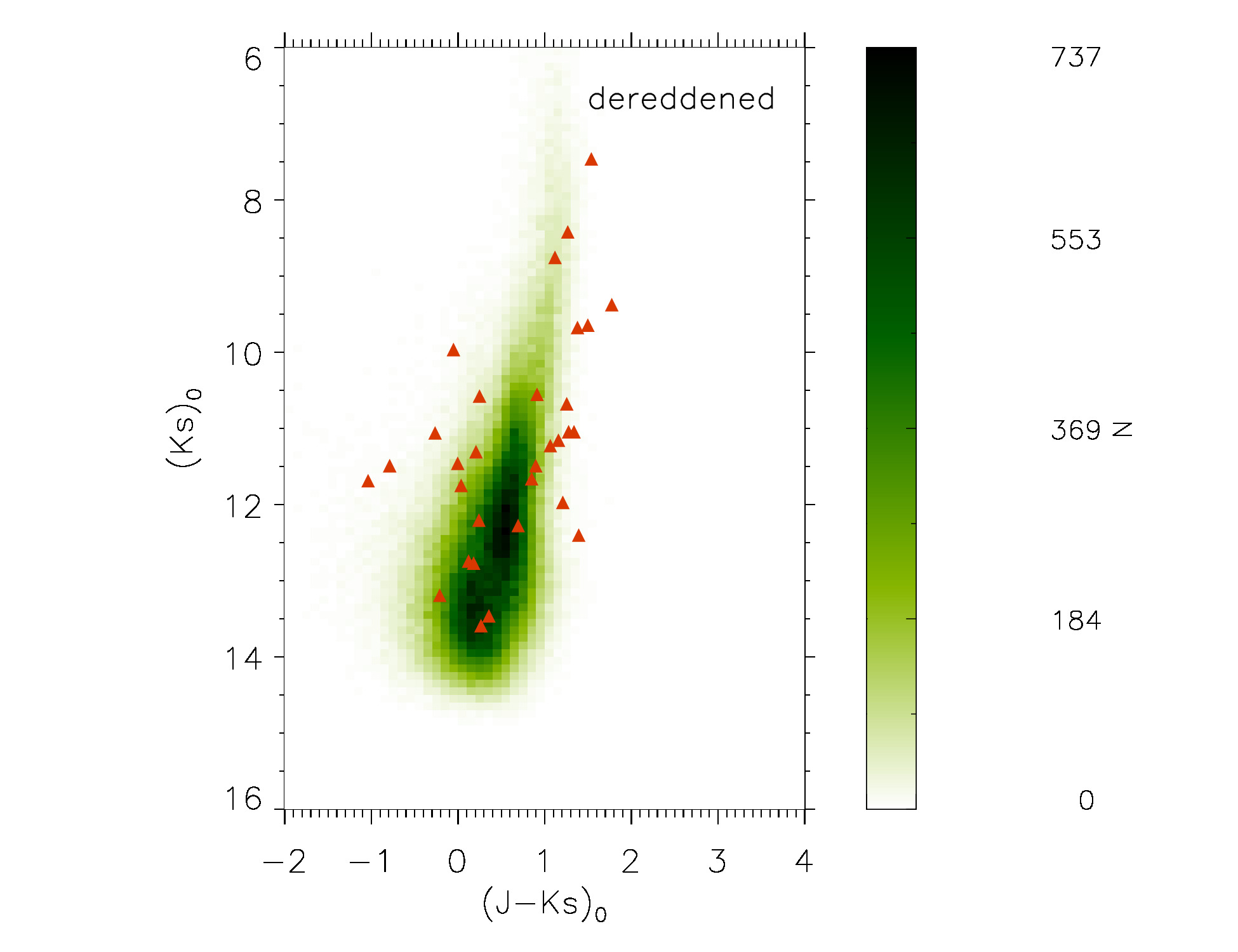}
   \caption{
   A star cluster in Mercer et al. (2005) discovered using the
   GLIMPSE point-source catalog. \emph{Top-left}, cutout of a GLIMPSE mosaic
   image in a square field with a side of
   $0.5^{\circ}$. \emph{Top-right}, extinction map for the same area
   which has combined data from tiles $d061$ and $d062$.   
   \emph{Bottom-left}, the raw CMD for the selected field where stars within $1\farcm2$ of the
   cluster centre appear as red triangles and the respective dereddened CMD (\emph{bottom-right}).
}
              \label{fig:glimpse}%
\end{figure*}

Lastly, to further check that the RJCE method performs satisfactorily when tackling a suite of diverse objectives, the approach was applied to deredden a star cluster. 
Figure \ref{fig:glimpse} shows a cluster located at Galactic coordinates l=328.3, b=-0.60 deg identified by Mercer et al. (2005) using GLIMPSE data. We show the cluster near-IR color image centered in a 0.5$^{\circ}$ square field, and the respective extinction map as derived from tiles d061 and d062.
The dereddened CMD of the same area is likewise shown, where stars in a $1\farcm2$ circle diameter of the cluster centre are highlighted. The cluster stars in the CMD belong primarily to the main sequence, and those in more evolved stages are either foreground targets or are bright enough to be visible despite foreground dust.  We expect that in the future the approach detailed here can be included in an automated detection routine seeking overdensities in the VVV data, throughout the Southern Galactic disk.

\section{Conclusions and Future Work}
 A high-resolution and deep RJCE map is presented here for the
 Southern Galactic disk.  
 The map is based on deep NIR VVV data combined with GLIMPSE, and complemented by 2MASS observations.
 The maps are consistent with results in the literature, where the main differences partly arise owing to selection effects based on the techniques and data employed.
 The differences highlight that differential extinction can be found 
 in regions as small as 2'/pixel resolution.  Nevertheless, the commonalities found among the maps is reassuring given their semi-independent nature.   
 In the future we plan to improve our maps by replacing the CASU aperture photometry with VVV DoPHOT PSF photometry (Alonso-Garc\'ia 2018). The new photometry will bolster the overall robustness of $A_{Ks}$ by increasing the number of stars per pixel, and enable the maps to rely on VVV NIR data rather than a VVV-2MASS combination. 
 At the same time, we will explore the effects of the variation of the extinction law, which has been set to a constant value in this work, by combining our new results with existing reddening maps obtained with other techniques.

\section*{Acknowledgements}
We would like to thank the anonymous referee for comments and advice that have greatly improved the clarity of the paper. 
M.S. acknowledges support from Becas Chile de Postdoctorado en el 
Extranjero project  74150088.
We gratefully acknowledge data from the ESO Public Survey program ID 179.B-2002 taken with the VISTA telescope, and products from the Cambridge Astronomical Survey Unit (CASU). 
D.M., J.B., R.K. and J.A.G. gratefully acknowledge support provided by the Ministry for the Economy, Development and Tourism, Programa Iniciativa Cientifica Milenio grant IC120009, awarded to the Millennium Institute of Astrophysics (MAS). D.M. also acknowledges support from project Fondecyt No. 1170121. D.G. and D.M. gratefully acknowledge support provided by the BASAL Center for Astrophysics and Associated Technologies (CATA) through grant AFB-170002. D.G. also acknowledges financial support from the Direccion de Investigacion y Desarrollo de la Universidad de La Serena through the Programa de Incentivo a la Investigacion de Academicos(PIA-DIDULS).
M.H. gratefully acknowledges support from the BASAL Center for Astrophysics and Associated Technologies (CATA) through grant PFB-06 and Comité Mixto ESO-Gobierno de Chile.
R.K.S. acknowledges support from CNPq/Brazil through projects 308968/2016-6 and 421687/2016-9.
CMB acknowledges support from FONDECYT regular project 1150060.
This material is based upon work supported in part by the
National Science Foundation under Grant No. 1066293 and the
hospitality of the Aspen Center for Physics. 
V.F. acknowledges support from CONICYT Astronomy Program-2015 Research Fellow
GEMINI-CONICYT (32RF0002).
F.M. gratefully acknowledges support from Programa DIDULS PT17145.
J.L.N.C. is also grateful for financial support received from the Programa
de Incentivo a la Investigaci\'on Acad\'emica de la Direcci\'on de
Investigaci\'on de la Universidad de La Serena (PIA-DIULS), Programa 
DIULS de Iniciaci\'on Cient\'ifica No. PI 15142. JLNC also
acknowledges the financial support from the GRANT PROGRAM 
N$^{\circ}$FA9550-15-1-0167 of the Southern Office of Aerospace Research
 and development (SOARD), a branch of the Air Force Office of the
 Scientific Researchs International Office  of the United States (AFOSR/IO).












\bsp	
\label{lastpage}
\end{document}